\documentclass[journal]{IEEEtran}
\ifCLASSINFOpdf
\else
\fi

\usepackage{booktabs} 
\usepackage{subfig}
\usepackage{graphicx}
\usepackage{algorithm}
\usepackage{algpseudocode}%
\usepackage{xcolor}
\usepackage{amsmath}
\newcommand{\etal}{\textit{et al}. }
\usepackage{mathtools}
\usepackage{multirow}
\usepackage{graphics}
\usepackage{enumitem}
\usepackage{CJKutf8}
\usepackage{array}
\usepackage{diagbox}
\usepackage{hyperref}
\usepackage[utf8]{inputenc}
\usepackage{ragged2e}
\usepackage{amsmath,amssymb}

\usepackage{xcolor,cite,etoolbox}
\definecolor{mypurple}{rgb}{0.4392, 0.1882, 0.6275}
\definecolor{mygreen}{rgb}{0, 0.6902, 0.3137}
\makeatletter 
\pretocmd\@bibitem{\csname keycolor#1\endcsname}{}{\fail}
\newcommand\citecolor[1]{\@namedef{keycolor#1}{}}
\makeatother
\begin{document}
%
\title{Unsupervised Time-Aware Sampling Network \\ with Deep Reinforcement Learning \\ for EEG-Based Emotion Recognition}
%
%
%
\author{\IEEEauthorblockN{Yongtao~Zhang\textsuperscript{a,b,}\IEEEauthorrefmark{10},
Yue~Pan\textsuperscript{a,b,}\IEEEauthorrefmark{2},
Yulin~Zhang\textsuperscript{a,b,}\IEEEauthorrefmark{3},
Linling~Li\textsuperscript{a,b,}\IEEEauthorrefmark{4},
Li~Zhang\textsuperscript{a,b,}\IEEEauthorrefmark{5},\\
Gan~Huang\textsuperscript{a,b,}\IEEEauthorrefmark{6},
Zhen~Liang\textsuperscript{a,b,}\IEEEauthorrefmark{9}\textsuperscript{,*} and Zhiguo~Zhang\textsuperscript{c,d,e,}\IEEEauthorrefmark{8}\textsuperscript{,*}}\\
\medskip
\IEEEauthorblockA{\small{{\textsuperscript{a}School of Biomedical Engineering, Health Science Center, Shenzhen University, Shenzhen, China\\
\textsuperscript{b}Guangdong Provincial Key Laboratory of Biomedical Measurements and Ultrasound Imaging, Shenzhen, China\\
\textsuperscript{c}Institute of Computing and Intelligence, Harbin Institute of Technology, Shenzhen, China\\
\textsuperscript{d}Marshall Laboratory of Biomedical Engineering, Shenzhen, China\\
\textsuperscript{e}Peng Cheng Laboratory, Shenzhen, China\\
\medskip
Email: \IEEEauthorrefmark{10}zhangyongtao2021@email.szu.edu.cn,
\IEEEauthorrefmark{2}2019222008@email.szu.edu.cn,
\IEEEauthorrefmark{3}2020222001@email.szu.edu.cn,
\IEEEauthorrefmark{4}lilinling@szu.edu.cn,
\IEEEauthorrefmark{5}lzhang@szu.edu.cn,
\IEEEauthorrefmark{6}huanggan@szu.edu.cn,
\IEEEauthorrefmark{9}janezliang@szu.edu.cn,
\IEEEauthorrefmark{8}zhiguozhang@hit.edu.cn}}}}

\maketitle

\begin{abstract}
Recognizing human emotions from complex, multivariate, and non-stationary electroencephalography (EEG) time series is essential in affective brain-computer interface. However, because continuous labeling of ever-changing emotional states is not feasible in practice, existing methods can only assign a fixed label to all EEG timepoints in a continuous emotion-evoking trial, which overlooks the highly dynamic emotional states and highly non-stationary EEG signals. To solve the problems of high reliance on fixed labels and ignorance of time-changing information, in this paper we propose a time-aware sampling network (\textbf{TAS-Net}) using deep reinforcement learning (DRL) for unsupervised emotion recognition, which is able to detect key emotion fragments and disregard irrelevant and misleading parts. Specifically, we formulate the process of mining key emotion fragments from EEG time series as a Markov decision process and train a time-aware agent through DRL without label information. First, the time-aware agent takes deep features from a feature extractor as input and generates sample-wise importance scores reflecting the emotion-related information each sample contains. Then, based on the obtained sample-wise importance scores, our method preserves top-\emph{X} continuous EEG fragments with relevant emotion and discards the rest. Finally, we treat these continuous fragments as key emotion fragments and feed them into a hypergraph decoding model for unsupervised clustering. Extensive experiments are conducted on three public datasets (SEED, DEAP, and MAHNOB-HCI) for emotion recognition using leave-one-subject-out cross-validation, and the results demonstrate the superiority of the proposed method against previous unsupervised emotion recognition methods. The proposed TAS-Net has great potential in achieving a more practical and accurate affective brain-computer interface in a dynamic and label-free circumstance. The source code is made available at \textit{https://github.com/infinite-tao/TAS-Net}.
\end{abstract}

\begin{IEEEkeywords}
Electroencephalography; Affective Brain-Computer Interface; Emotion Recognition; Deep Reinforcement Learning; Unsupervised Learning.
\end{IEEEkeywords}

\footnotetext[1]{\hspace{1mm}Corresponding author: Zhen Liang and Zhiguo Zhang.}

%
\IEEEpeerreviewmaketitle

\section{Introduction}
\label{sec:introduction}
\IEEEPARstart{H}{uman} emotion is a complex and dynamic process involving different levels of processing and integration \cite{cowen2017self,horikawa2020neural}. In emotion-related studies, how to accurately describe and further recognize human emotions has been a critical issue in the last decade. Recently, the developments in using electroencephalography (EEG) signals for emotion recognition have gained increasing attention from researchers with a various of backgrounds \cite{huang2014novel,liu2017real,song2018eeg,li2022eeg}, because EEG could provide a more direct and objective clue to understand and estimate emotional states. The current EEG-based emotion recognition studies are sample-based emotion recognition, which treats the EEG samples in a continuous data collection equally and assigns the fixed emotional label (trial-based groundtruth), without taking into account emotion dynamics. As a consequence, the highly viable emotional states and associated EEG signals are not explored comprehensively, which hinders our understanding of emotion dynamics and decreases the accuracy of emotion recognition. To solve these problems, a novel EEG-based emotion recognition approach is proposed in this paper to unsupervisedly and adaptively determine key emotion fragments from an EEG trial which would be then used for emotion recognition.

Currently, many existing EEG-based emotion recognition models have been built via supervised machine learning, which can roughly fall into two categories. (1) Traditional machine learning-based methods. Based on the commonly used handcrafted EEG features, such as power spectral density (PSD) \cite{alsolamy2016emotion}, differential entropy (DE) \cite{duan2013differential}, differential asymmetry (DASM) \cite{duan2012eeg}, rational asymmetry (RASM) \cite{lin2010eeg}, and differential caudality (DCAU) \cite{zheng2015investigating}, machine learning-based classifiers are built for emotion recognition \cite{duan2013differential,liu2013real,bahari2013eeg}. For example, Alsolamy \textit{et al.} \cite{alsolamy2016emotion} extracted the PSD features at different frequency bands and used a support vector machine (SVM) as a classifier to predict emotions during listening to Quran. To enhance emotion recognition performance, Atkinson \textit{et al.} \cite{atkinson2016improving} introduced to use the minimum-Redundancy-Maximum-Relevance (mRMR) method to select the most emotion-related handcrafted features for modeling. However, due to the unstable and weak performance of the traditional machine-learning methods, more and more EEG-based emotion recognition models are developed by using (2) deep learning-based methods. EEG-based emotion recognition models with deep learning are able to automatically extract discriminant features and identify emotional states from EEG signals in an end-to-end manner \cite{pan2010domain,liu20213dcann,zheng2015investigating,li2019domain}. For example, Li \textit{et al.} \cite{li2018novel} proposed a bi-hemispheres domain adversarial neural network (BiDANN) based on the asymmetry between the left and right hemispheres of the brain, which made the data representation easy for emotion recognition by separately mapping the EEG data of the left and right hemispheres into discriminant feature spaces. Song \textit{et al.} \cite{song2018eeg} proposed a dynamical graph convolutional neural network (DGCNN) with five types of handcrafted features as input, which dynamically learns the intrinsic relationships between different EEG channels represented by an adjacency matrix. Zhong \textit{et al.} \cite{zhong2020eeg} proposed an EEG-based regularized graph neural network (RGNN) for performance enhancement. Li \textit{et al.} \cite{li2019regional} proposed to extract discriminative spatial-temporal EEG features via a region-to-global feature learning process. Tao \textit{et al.} \cite{tao2020eeg} developed an attention-based convolutional recurrent neural network (ACRNN) to adaptive estimate the importance of temporal and spatial information in EEG signals using the channel attention mechanism and self-attention mechanism. Compared with traditional machine learning-based methods, deep learning-based methods have achieved great success in EEG-based emotion recognition due to their powerful data representation ability.

Most of the existing deep learning methods for EEG-based emotion recognition require a large number of training samples with emotional labels. However, it is unrealistic to collect a large number of EEG signals from different participants and manually annotate each sample based on emotional labels. To increase the sample size, previous studies usually divided each EEG trial into samples with a fixed length of 1s and annotated these samples with the fixed emotional label (trial-based groundtruth), which leads to the failure of neural networks to learn the true data distribution \cite{zheng2015investigating,zheng2016multichannel,li2019domain}. In our previous work \cite{liang2019unsupervised}\cite{liang2021eegfusenet}, we proposed unsupervised learning-based methods to eliminate the model's reliance on labels. In particular, Liang \etal \cite{liang2021eegfusenet} developed a hybrid deep convolutional recurrent generative adversarial network named EEGFuseNet, which automatically characterized spatial and temporal dynamics from EEG signals in a self-learning paradigm and realized a possible way for unsupervised EEG feature learning in the affective brain-computer interface applications. However, the emotion dynamics in a trial were still underestimated. In a continuous emotion-evoking process, the elicited emotions are in a state of flux and the simultaneously collected EEG signals should not be assigned the fixed emotional label.

\begin{figure}[t]
\begin{center}
\includegraphics[width=0.5\textwidth]{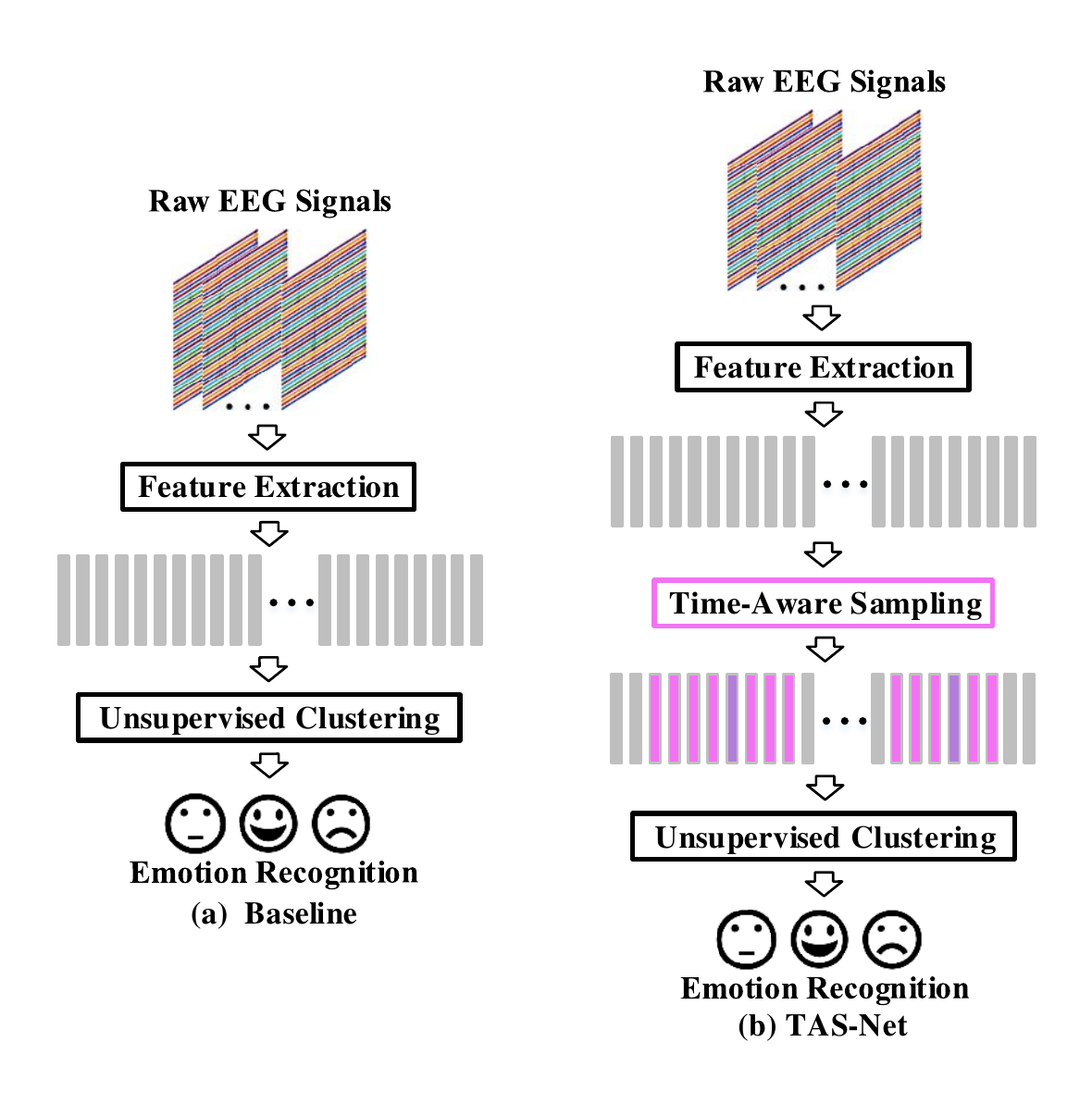}
\end{center}
\caption{The flowcharts of (a) the baseline method and (b) the proposed TAS-Net for unsupervised emotion recognition, respectively. In the proposed TAS-Net, the key emotion fragments (purple and light-purple parts) are firstly selected via a DRL-based time-aware sampling method, and then only the selected key emotion fragments are used for emotion recognition, as opposed to using all the samples in the baseline method.}
\label{fig:flowchart}
\end{figure}

With the rapid development of deep learning technology, deep reinforcement learning (DRL) has been applied in many real-world scenarios, such as games \cite{lample2017playing}, robotics \cite{wang2016does}, natural language processing \cite{he2015deep}, visual understanding \cite{yarats2020image}, and neural architecture search \cite{zoph2016neural}. Unlike the conventional machine learning method, DRL learns through the reward signals of actions. In other words, a reasonable reward function is essential for a specific task based on DRL. Mnih \etal \cite{mnih2013playing} successfully approximated value function with a deep CNN, and enabled their agent to beat a human expert in several Atari games. Later on, researchers start to apply DRL algorithms to time-series processing such as EEG analysis \cite{zhang2018fuzzy,li2021eeg}. For example, Zhang \etal \cite{zhang2018fuzzy} introduced DRL to aggregate local spatio-temporal and global temporal information from EEG signals for movement intention detection. Li \textit{et al.} \cite{li2021eeg} developed a neural architecture search framework by using DRL to automatically design network architectures and learn discriminative EEG features for emotion recognition. However, these methods also ignore the emotion dynamics in a trial. Inspired by the video summarization method presented in \cite{zhou2018deep} which detects key video frames to largely enhance the model efficiency for video processing, we introduce a novel time-aware sampling method to automatically detect key emotion fragments from the time-series EEG signals in order to fully consider the emotion dynamics in a trial. In comparison to \cite{zhou2018deep}, we make further improvements based on the characteristics of emotion occurrence, including agent design, reward improvement, and detection strategy. In other words, for each trial, only the EEG data located at the selected key emotion fragments are retained for further unsupervised emotion recognition, and other less informative data are discarded. Here, we take "EEGFuseNet + Unsupervised Clustering" as the baseline method, and highlight the flowchart difference between the baseline method and the proposed method (TAS-Net) in Fig. 1.

TAS-Net attempts to detect key emotion fragments in an unsupervised manner and further improves the model performance on EEG-based emotion recognition. First, an unsupervised deep feature extractor is employed to extract deep spatial and temporal features from raw EEG signals. Second, considering the non-stationary emotion in EEG time series, we propose a DRL-based time-aware sampling method to detect the most informative key emotion fragments from a trial. Here, a new time-aware agent that can better capture temporal dependencies is developed. It consists of a local time-based graph convolution network (GCN) and a global time-based bidirectional gated recurrent unit (BiGRU). Third, an unsupervised hypergraph decoding model is presented to categorize emotions based on the detected key emotion fragments. In summary, this work has three main contributions:
\begin{itemize}
\item We propose a novel unsupervised emotion recognition method based on deep reinforcement learning, which formulates the process of mining key emotion fragments from EEG time series as a Markov decision process. To the best of our knowledge, we are first to utilize deep reinforcement learning to locate informative emotion fragments for better unsupervised emotion recognition.
\item We develop a novel time-aware agent, which consists of local time-based GCNs to learn the intrinsic relationship between adjacent segments and a global time-based BiGRU to capture long-term dependencies in EEG time series.
\item We conduct comprehensive experiments on the emotion recognition task with the SEED, DEAP, and MAHNOB-HCI datasets. The proposed method not only achieves better performance than other unsupervised learning methods on the three public datasets, but is also comparable to some supervised learning methods under the same validation strategy. Additionally, compared to the baseline method, the performance using the proposed method is significantly improved.
\end{itemize}

\begin{figure*}
\begin{center}
\includegraphics[width=1\textwidth]{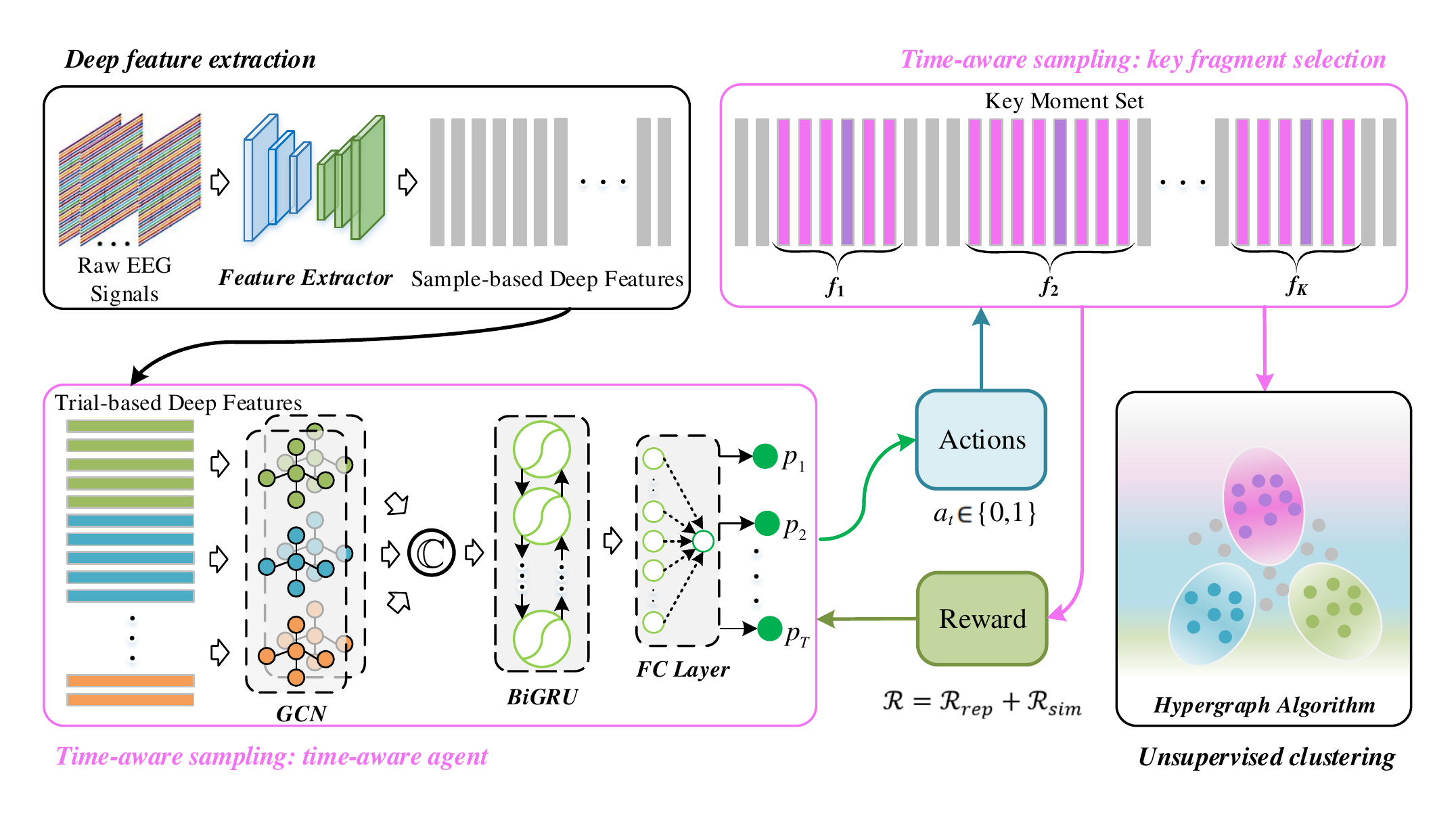}
\end{center}
\caption{The framework of the proposed TAS-Net, which includes \textit{deep feature extraction}, \textit{time-aware sampling}, and \textit{unsupervised clustering}. GCN: graph convolution network; \large{\textcircled{\normalsize{$\mathbb{C}$}}}: \normalsize{concatenation; BiGRU: bidirectional gated recurrent unit; FC Layer: fully connected layer; $p_{t}$: action probability score; $f_{k}$: the detected key emotion fragments. Here, the detected key emotion fragments are composed of the detected key moments (dark purple) and the emotion offsets (light purple) calculated according to the nature of "short-term continuity" in human emotions. The final detected key emotion fragments are composed of both dark and light purple parts.}}
\label{fig:DRL}
\end{figure*}

\section{Methodology}
\label{sec:methodology}
\subsection{Overview}
The framework of the proposed TAS-Net is shown in Fig. \ref{fig:DRL}, in which three parts are included: \textbf{(1) deep feature extraction:} extract sample-based deep EEG features; \textbf{(2) time-aware sampling}: detect trial-based informative emotion fragments; and \textbf{(3) unsupervised clustering:} realize unsupervised emotion recognition using EEG signals. Here, the collected raw EEG signals in one trial are first segmented into a number of samples with a fixed length of 1s and input to the deep feature extractor of the baseline method to extract sample-based deep EEG features. Second, the extracted sample-based deep EEG features are formed into trial-based representation and input to the time-aware agent (the first part in the time-aware sampling) to measure the contribution of each sample to the emotion dynamics in one trial. The output is a sequence of probabilities indicating how informative each sample is in referring to emotion dynamics in a trial. The EEG samples with high probabilities are then selected as the informative emotion samples in a trial. Considering that human emotion has "short-term continuity", we further extract top-\emph{X} continuous fragments centered on the selected informative emotion samples through the key fragment selection (the second part in the time-aware sampling). Finally, the extracted informative emotional fragments are used for emotion recognition using the unsupervised clustering method. More details about each module in the proposed TAS-Net are presented below.

\subsection{Deep Feature Extraction}
\label{sec:feature extraction}
An unsupervised deep feature extraction network (EEGFuseNet) was proposed \cite{liang2021eegfusenet} to efficiently extract reliable features from high-dimensional raw EEG signals. Therefore, we employ the EEGFuseNet as our feature extractor to characterize sample-based deep EEG features. Specifically, EEGFuseNet was a hybrid deep encoder-decoder network architecture, which integrated a convolutional neural network (CNN), recurrent neural network (RNN), and generative adversarial network (GAN) in a smart and efficient manner. In the CNN-RNN based deep encoder-decoder network, the spatial and temporal dynamics in a time-series EEG data were efficiently characterized to represent the complex relationships of brain signals at different brain locations and at different time points. Based on the extracted spatial information by CNN, RNN was then adopted to measure the temporal information by exploring the feature relationships at adjacent time points. To further benefit high-quality feature characterization, a discriminator was incorporated to enhance the generator (CNN-RNN based)'s performance. EEGFuseNet is a self-learning network without any reliance on labeled samples, which is suitable to solve unsupervised EEG processing.

\subsection{Time-Aware Sampling}
\label{sec:sampling}
In a continuous emotion-evoking process, not all the timepoint are equally important to the final elicited emotions. In other words, not all of the simultaneously collected EEG data in a trial should be treated equally using the fixed emotional label. Only a small percentage of EEG samples in a trial may contain significant and relevant emotion information, whereas the remaining data may represent less emotion information that could be discarded in emotion recognition. In this section, we present the proposed time-aware sampling method that adaptively highlights the most informative emotion fragments from each trial, without any need for label information.

Following the deep feature extraction, the time-aware agent in the proposed time-aware sampling approach interprets the key moment detection task as a Markov decision process. During the training phase, the time-aware agent interacts with an environment that provides rewards and takes actions $ a_{t} $ on whether a sample should be selected into the key moment set $ \mathcal{S} $ by maximizing the total expected reward $ \mathcal{R} $. In the inference stage, we retain the top-\emph{X} continuous fragments centered on key emotion fragments for unsupervised clustering. The state, agent, action, and reward of the Markov decision process are described below.

\textbf{State}. The state $ s $ is the extracted deep features in the \textit{deep feature extraction} part.

\textbf{Agent}. Considering that the human emotion has not only ”short-term continuity” but also ”long-term similarity”, the time-aware agent is designed to consist of two components: a local time-based GCN and a global time-based BiGRU topped with a FC layer. Given a sequence of trial-based deep features $ \{ {s_{t}}\}_{t=1}^{T} $ calculated by the feature extractor, the time-aware agent first divides the EEG samples in a trial into $ M $ successive segments ($ m\in \{1,...,M\} $), which can be explained as follows:
\begin{equation}
    \label{Eq:segment}
    C_{m}=
    \{ s_{i}: i = \left(m-1\right) \cdot L+1,..., min \{T, m \cdot L \} \},
\end{equation}
where $ L $ denotes the time length of a successive segment, and $T$ is the length of one trial. Then, the local time-based GCN takes these successive segments as input and estimates the local temporal features in parallel. After that, we concatenate these segment-based local temporal features in the time dimension and input them into the global time-based BiGRU to predict an action probability score ${p_{t}}$ for each sample. The action probability score is defined as:
\begin{equation}
    \label{Eq:probability}
    p_{t}= \sigma \left(W h_{t} \right),
\end{equation}
where $\sigma$ represents the sigmoid function, $W$ is the trainable parameter of the FC layer, and $h_{t} = \{ h_{t}^{GCN}, h_{t}^{BiGRU} \}$ denotes the time-aware agent produces hidden state for $t$-th sample.

\textbf{Action}. The designed time-aware agent further takes actions to select the key emotion fragments according to the obtained action probability score, and the action $ \alpha_{t} $ is given as:
\begin{equation}
    \label{Eq:action}
    a_{t} \sim {\rm Bernoulli} \left(p_{t}\right),
\end{equation} 
where $ a_{t} \in \{0,1\} $ indicates whether $t$-th sample in a trial should be selected in the key moment set or not. Therefore, the key moment set $ \mathcal{S} $ can be defined as:
\begin{equation}
    \label{Eq:keyEmo}
    \mathcal{S} = \{ s_{y_{i}} \mid a_{y_{i}} = 1, i=1,2,...\},
\end{equation}
where $ y_{i} $ represents the indices of the selected samples.

\textbf{Reward}. To optimize the key moment selection of the time-aware agent and guarantee all the informative and discriminant points in a trial are covered, we introduce two reward functions to optimize the training process:
\begin{equation}
    \label{Eq:reward}
    \mathcal{R} = \mathcal{R}_{rep} + \mathcal{R}_{sim},
\end{equation}
where $\mathcal{R}_{rep}$ and $\mathcal{R}_{sim}$ denote the representativeness reward and the similarity reward, respectively. $\mathcal{R}_{rep}$ measures how well the selected key moment set $ \mathcal{S} $ represents the original trial data, given as
\begin{equation}
    \label{Eq:rewardRep}
    \mathcal{R}_{rep} = {\rm exp}\left( -\frac{1}{T}\sum_{t=1}^{T}\min_{t^{'}\in\mathcal{Y}}\left\|s_{t}-s_{t^{'}}\right\|_2\right),
\end{equation}
where $ \mathcal{Y} = \{y_{i}\mid a_{y_{i}} = 1, i=1,2,..., |\mathcal{Y}|\}. $ Through maximizing $\mathcal{R}_{rep}$, the model tends to preserve the key emotion fragments that could represent the overall temporal information of the input trial data. On the other hand, to minimize the information redundancy on the selected key emotion fragments, we also introduce the similarity reward ($\mathcal{R}_{sim}$) to estimate the similarities among those selected key moment set $ \mathcal{S} $, given as:
\begin{equation}
    \label{Eq:rewardSim}
    \mathcal{R}_{sim} = \frac{1}{\mid \mathcal{Y} \mid \left(\mid\mathcal{Y}\mid-1\right)}\sum_{t\in\mathcal{Y}}\sum_{t^{'}\in \mathcal{Y},t^{'}\neq t} sim\left(s_{t},s_{t^{'}}\right),
\end{equation}
where $ sim\left(\cdot,\cdot\right) $ is the cosine similarity of two vectors.

\subsubsection{Training with Policy Gradient}
Algorithm 1 shows the training details of the proposed time-aware agent. The goal of the time-aware agent is to learn a policy function $\pi_{\theta}$ with parameters $\theta$ by maximizing the expected reward:
\begin{equation}
    \label{Eq:theta}
    J\left(\theta\right) = \mathbb{E}_{p_{\theta}\left(a_{1:T}\right)}[\mathcal{R}],
\end{equation}
where $p_{\theta}\left(a_{1:T}\right)$ represents the probability distribution over possible action sequences, and $\mathcal{R}$ is calculated by Eq. (5). According to the definition given above, the obtained action sequences could consist of different selection choices on the key emotion fragments. There are 2 actions for each segment and the exponential $2^{T}$ is computationally infeasible for the time-aware agent training. Thus, we employ the policy gradient method to efficiently compute the continuous action space. 

Inspired from the REINFORCE algorithm proposed in \cite{williams1992simple}, we compute the derivative of the expected reward $J\left(\theta\right)$ w.r.t. the parameters $\theta$ as:
\begin{equation}
    \label{Eq:thetaPara}
\nabla_{\theta}J\left(\theta\right)=\mathbb{E}_{p_{\theta}\left(a_{1:T}\right)}[\mathcal{R}\sum_{t=1}^{T}\nabla_{\theta}{\rm log}\pi_{\theta}\left(a_{t} \mid h_{t}\right)],
\end{equation}
where $a_{t}$ is the action taken by the time-aware agent for $t$-th sample, and $h_{t}$ is the hidden state from the time-aware agent. As Eq. (9) involves the expectation over high dimensional action sequences, which is difficult to compute directly. Here, we approximate the gradient by running the agent for $ N $ episodes on the same trial data and then taking the average gradient as follows:
\begin{equation}
    \label{Eq:gradient}
    \nabla_{\theta}J\left(\theta\right)\approx\frac{1}{N}\sum_{n=1}^{N}\sum_{t=1}^{T}\mathcal{R}_{n}\nabla_{\theta}{\rm log}\pi_{\theta}\left(a_{t} \mid h_{t}\right),
\end{equation}
where $\mathcal{R}_{n}$ is the reward computed at the $n$-th episode. Although the gradient in Eq. (10) gives a direction for updating $\theta$, it may contain high variance and lead to poor network convergence. To tackle this issue, we normalize the reward by subtracting the reward using a constant baseline $b$. The gradient becomes:
\begin{equation}
    \label{Eq:gradient2}
    \nabla_{\theta}J\left(\theta\right)\approx\frac{1}{N}\sum_{n=1}^{N}\sum_{t=1}^{T}\left(\mathcal{R}_{n}-b\right)\nabla_{\theta}{\rm log}\pi_{\theta}\left(a_{t} \mid h_{t}\right),
\end{equation}
where $b$ is a moving average of rewards experienced so far for computational efficiency enhancement.

On the other hand, to avoid too many key moments selected for increasing the reward, we also introduce a regularization term ($ \mathcal{L}_{sampling} $) to limit the selection percentage for the key moment set. The regularization is defined as:
\begin{equation}
    \label{Eq:regularization}
    \mathcal{L}_{sampling}=\mid\mid\frac{1}{T}\sum_{t=1}^{T}p_{t}-\vartheta\mid\mid^{2},
\end{equation}
where $ \vartheta $ is a scalar representing the percentage of key emotion fragments to be selected. Finally, we optimize the time-aware agent through a joint loss of the reward functions and regularization:
\begin{equation}
    \label{Eq:finalLoss}
    \mathcal{L}_{total}=\beta\mathcal{L}_{sampling}-\mathcal{R},
\end{equation}
where $ \beta $ denotes a balance factor of the two terms.

\begin{table}
\label{tab:Algo1}
\begin{tabular}{l}
\toprule
\textbf{Algorithm 1} Time-aware agent training \\
\midrule
\textbf{Input:} Deep feature sequence $ \{ {s_{t}}\}_{t=1}^{T} $ from training set \\
\textbf{Output:} Parameters $\theta$ of the time-aware agent\\
1: \textbf{Initialize} $\theta$ with small random values\\
2: \textbf{for} $ i \leftarrow 1,2,..., \xi $ \textbf{do}\\
3: \quad \ \ Input a deep feature sequence $ \{ {s_{t}}\}_{t=1}^{T} $ from training set\\
4: \quad \ \ Compute action probabilities $\{ {p_{t}}\}_{t=1}^{T}$\\
5: \quad \ \ \textbf{for} $ episode \leftarrow 1,2,..., \textit{N} $ \textbf{do}\\
6: \qquad \quad Sample $ \{{\textit{a}_{t}}\}_{t=1}^{T} $ from policy $ \pi_{\theta} $ \\
7: \qquad \quad \ Compute reward $ \mathcal{R}_{n} $ \\
8: \quad \ \ \textbf{end for}\\
9: \quad \ \ Update $\theta$ with gradient $\nabla_{\theta}J\left(\theta\right)$ and $\nabla_{\theta}L_{sampling}$\\
10:\quad \   Update the moving average of rewards $ b $ for the corresponding \\
\quad \quad \; deep feature sequence. \\
11: \textbf{end for}\\
12: \textbf{return} $\theta$\\
\bottomrule
\end{tabular}
\end{table}

\begin{table}
\label{tab:Algo2}
\color{black}
\begin{tabular}{l}
\toprule
\textbf{Algorithm 2} Time-aware agent testing \\
\midrule
\textbf{Input:} Deep feature sequence $ \{{s_{t}}\}_{t=1}^{T} $ from testing set \\
\textbf{Output:} Action probabilities $ \{{p_{t}}\}_{t=1}^{T} $\\
1: \textbf{Initialize} $\theta$ from the trained time-aware agent\\
2: Input a deep feature sequence $ \{ {s_{t}}\}_{t=1}^{T} $ \\
3: Compute importance scores $\{ {p_{t}}\}_{t=1}^{T}$ through policy $ \pi_{\theta} $ \\
4: Sort importance scores in descending order \\
5: Select the top-$X$ continuous fragments according to Eq. (14)\\
6: Save the top-$X$ continuous fragments for unsupervised clustering\\
7: \textbf{return $ \{{p_{t}}\}_{t=1}^{T} $} \\
\bottomrule
\end{tabular}
\end{table}

\subsubsection{Key fragment selection in inference}
Algorithm 2 shows the key fragment selection part in the proposed TAS-Net method. Given a deep feature sequence $ \{ {s_{t}}\}_{t=1}^{T} $ from a testing set, we feed it into the trained time-aware sampling model $ \pi $ to predict sample-level probabilities as importance scores. We first sort the sample-wise importance scores in descending order and select top-$ X $ points as the key moments. Considering the nature of "short-term continuity" in human emotions, we further use the selected key moments to form top-$ X $ emotion offsets which are centered around the selected key moments. The emotion offsets are defined as:
\begin{equation}
    \label{Eq:fragment}
    f_{x}=\{ s_{\lceil o_{x}^{l} \rceil}, s_{\lceil o_{x}^{l} \rceil +1},..., s_{c_{x}},..., s_{\lfloor o_{x}^{r} \rfloor} \}, x \in \{1,2,...,X\},
\end{equation}
where $ s_{c_{x}} $ is the key moment of $ f_{x} $, and $ s_{c_{x}} \in S $. $ \lceil \cdot \rceil $ and $ \lfloor \cdot \rfloor $ represent the ceiling and floor operators, respectively. $ o_{x}^{l} $ and $ o_{x}^{r} $ denote two absolute offsets, which are formulated as:
\begin{equation}
    \label{Eq:centroid_down}
    o_{x}^{l}=c_{x}-\tilde{o}_{x}^{l} \cdot l_{max},
\end{equation}
\begin{equation}
    \label{Eq:centroid_up}
 o_{x}^{r}=c_{x}+\tilde{o}_{x}^{r} \cdot r_{max}.
\end{equation}
where $ \tilde{o}_{x}^{l} $ and $ \tilde{o}_{x}^{r} $ are relative offset coefficients, which are equal to the importance score $ p_{c_{x}}$ at $ c_{x} $-th key moment. $ l_{max} \in \mathbb{Z} \geq 0 $ is the maximum absolute offset to the left, and $ r_{max} \in \mathbb{Z} \geq 0 $ is the maximum absolute offset to the right. Finally, we obtain top-$X$ key emotion fragments for unsupervised emotion recognition.

\subsection{Unsupervised Clustering Model}
For a fair comparison with the baseline method, we also employ the hypergraph structure \cite{zhou2006learning} to perform unsupervised emotion recognition. Hypergraph is widely regarded as an effective method to describe complex hidden data structures, so it has natural advantages in decoding EEG signals and completing emotion classification. In a hypergraph, one hyperedge could connect more than two nodes and reveal more complex hidden structures than a pairwise connection in a simple graph (one edge can only connect two nodes). In the implementation, we formulate a hypergraph as $ \mathcal{G} = \left( \mathcal{V}, \mathcal{E}, w \right) $, where $ \mathcal{V} $ refers to vertices indicating the trial-based EEG representation obtained in the proposed time-aware sampling method and $ \mathcal{E} $ is the constructed hyperedge information based on the $ \mathcal{V} $. The unsupervised emotion recognition is realized by calculating the hypergraph Laplacian of the constructed hypergraph ($ \mathcal{G} $) and solving it in an optimal eigenspace.

\begin{figure}[t]
\begin{center}
\includegraphics[width=0.5\textwidth]{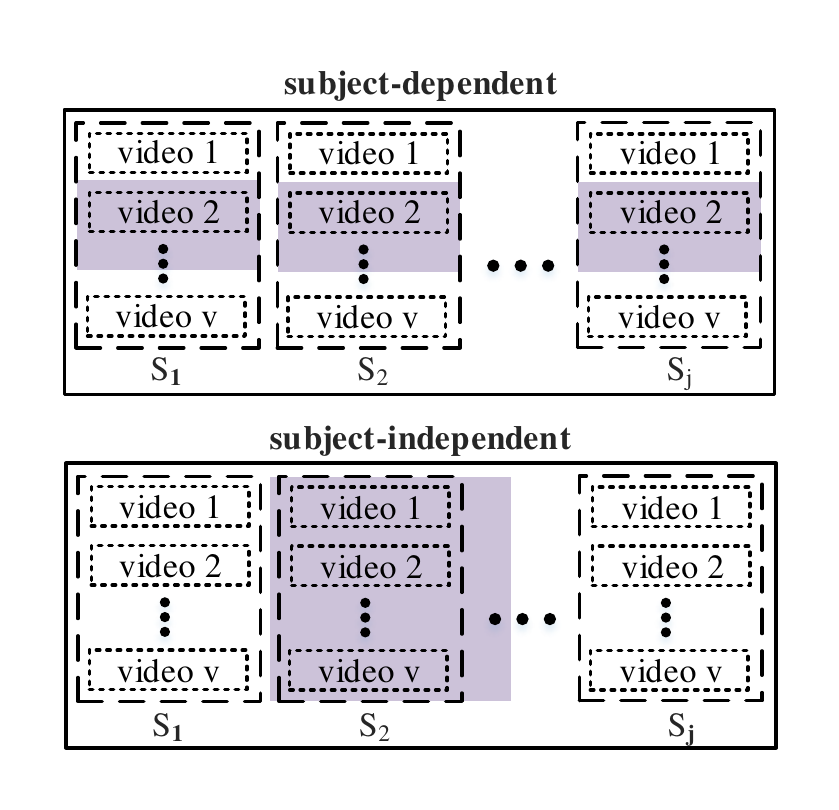}
\end{center}
\caption{Two validation strategies for EEG-based emotion recognition. The subject-dependent validation strategy divides data sets (purple represents the test set and the rest is the training set) within the subject, which can be subdivided into two categories: video-level K-fold cross-validation (CV) and video-level leave-one-out cross-validation (LOOCV). The subject-independent validation strategy divides data sets (purple represents the test set and the rest is the training set) among subjects, which can be subdivided into two categories: subject-level K-fold CV and subject-level LOOCV. $\rm{v}$ refers to the number of videos used in the experiment, and $ \rm {S_{j}} $ represents j-th subject.}
\label{fig:VS}
\end{figure}

\section{Experimental Results}
\label{sec:experiment}
We conduct a comprehensive experiment to evaluate the effectiveness of the proposed TAS-Net in dealing with unsupervised emotion recognition using EEG signals. In this section, we would first describe the datasets used in our experiments, followed by an introduction to experiment settings and evaluation metrics. Then, we would report the experimental results. The promising results demonstrate the effectiveness of the proposed method.

\subsection{Datasets}
We perform experiments on three publicly available EEG datasets, including SEED \cite{zheng2015investigating}, DEAP\cite{koelstra2011deap}, and MAHNOB-HCI\cite{soleymani2011multimodal}. All these datasets have been widely used as benchmarks for evaluating EEG-based emotion recognition algorithms.

\begin{itemize}
\item \textbf{SEED.} The SEED dataset contains EEG signals of 15 subjects, which were recorded at a 62-channel setting according to the international 10-20 system \cite{kwak2002input} when each subject watched 15 film clips. These film clips can evoke specific target emotions (positive, neutral, or negative), and each film clip is about 4 minutes long.
\item \textbf{DEAP.} The DEAP dataset contains EEG signals of 32 subjects, which were recorded at a 32-channel setting according to the international 10-20 system when each subject watched 40 music videos. These music videos were all 60 seconds in duration, and each was given corresponding subjective feedback on different emotional dimensions (valence, arousal, dominance, and liking). The participants were asked to rate different emotion dimensions on a scale of 1 to 9 after watching each music video. To cross-compare with the baseline method and other methods, we use a fixed threshold of 5 for each emotion dimension to discretize the subjective feedback into two classes (low and high).
\item \textbf{MAHNOB-HCI.} The MAHNOB-HCI dataset contains EEG signals of 30 subjects, which were recorded at a 32-channel setting according to the international 10-20 system when each subject watched 20 video clips. These film clips were selected to evoke emotions and the subjective feedback was given using a score in the range of 1 to 9. In the model evaluation, a fixed threshold of 5 is used to discretize the subjective feedback into binaries for each emotion dimension (valence, arousal, dominance, and predictability).
\end{itemize}

\subsection{Experiment Settings and Evaluation Metrics}
The proposed TAS-Net is implemented in the PyTorch platform. Here, we set the $ L $ in Eq. (1) to 16, the $ \vartheta $ in Eq. (12) to 0.5, the $ \beta $ in Eq. (13) to 0.01, and the number of episodes $ N $ to 5. The dimension of the edge feature in the GCN is 32, and the dimension of the hidden state in the GRU cell is 256. The above hyper-parameters are set according to the relevant literature \cite{zhang2022eeg}. The proposed TAS-Net uses an Adam optimizer with an initial learning rate of $ 10^{\operatorname{-4}} $ and a weight decay of $ 10^{\operatorname{-5}} $. Besides, the training is stopped when it reaches a maximum number of epochs (100 in our case).

Inconsistent model validation strategies have been used in previous EEG-based emotion recognition methods, which can be divided into two categories, subject-dependent and subject-independent. Fig. \ref{fig:VS} shows the details of data partitioning for the above two validation strategies. The subject-level LOOCV strategy avoids information leakage, so it has lower recognition performance compared to other validation strategies. For a fair comparison with the baseline method, we mainly use the subject-level LOOCV strategy. In each validation round, only one subject's all data are used as the testing data, and the rest of the other subjects' data are treated as training data.

In our experiments, we employ two performance metrics (accuracy $ P_{acc} $ and F1-Score $ P_{f} $) to quantitatively evaluate the obtained emotion recognition results. $ P_{acc} $ is a common evaluation metric for the classification task, and $ P_{f} $ is more able to measure the classification performance of datasets with imbalanced classes. Note that all the results reported below are an average performance across all subjects in different datasets.

\begin{table}[]
\color{black}
\begin{center}
  \caption{Emotion recognition performance $\left(\%\right) $ on SEED dataset.}
  \label{tab:seedCompare}
  \scalebox{0.70}{
  \begin{tabular}{lccc}
   \midrule
    \toprule
      \ Methods & Classification Task & $P_{acc}$ & $P_{f}$ \\
       \midrule
           \ \textbf{\textit{Supervised}} & \multicolumn{2}{l}{\textbf{\textit{Subject-Dependent}}} & \multicolumn{1}{c}{\textbf{\textit{Video-level LOOCV}}}\\
     \midrule
     \ DBN \cite{zheng2015investigating} & 3-Class & 86.08 & - \\
     \ GSCCA \cite{zheng2016multichannel} & 3-Class & 82.96 & - \\
     \ BiDANN \cite{li2018novel} & 3-Class & 92.38 & - \\
      \ DGCNN \cite{song2018eeg} & 3-Class & 90.40 & - \\
     \ R2G-STNN \cite{li2019regional} & 3-Class & 93.38 & - \\
     \ RGNN \cite{zhong2020eeg} & 3-Class & 94.24 & - \\
        \midrule
       \ \textbf{\textit{Supervised with Transfer Learning}} & \multicolumn{2}{l}{\textbf{\textit{Subject-Independent}}} & \multicolumn{1}{c}{\textbf{\textit{Subject-level LOOCV}}}\\
        \midrule
        \ TCA \cite{pan2010domain} & 3-Class & 63.64 & - \\
         \ BiDANN \cite{li2018novel} & 3-Class & 83.28 & - \\
         \ DGCNN \cite{song2018eeg} & 3-Class & 79.95 & - \\
        \ R2G-STNN \cite{li2019regional} & 3-Class & 84.16 & - \\
        \ RGNN \cite{zhong2020eeg} & 3-Class & 85.30 & - \\
        \ JDA \cite{li2019domain} & 3-Class & 88.28 & - \\
        \midrule
         \ \textbf{\textit{Supervised without Transfer Learning}} & \multicolumn{2}{l}{\textbf{\textit{Subject-Independent}}} & \multicolumn{1}{c}{\textbf{\textit{Subject-level LOOCV}}}\\
          \midrule
         \ JDA \cite{li2019domain} (source domain only) & 3-Class & 58.23 & - \\
        \midrule
        \ \textbf{\textit{Unsupervised}} & \multicolumn{2}{l}{\textbf{\textit{Subject-Independent}}} & \multicolumn{1}{c}{\textbf{\textit{Subject-level LOOCV}}}\\
        \midrule
     \ \multirow{2}{*}{Baseline \cite{liang2021eegfusenet}} & {2-Class} & 80.83 & 82.03 \\
     \ & {3-Class} & 59.06 & -\\
     \ \multirow{2}{*}{\textbf{TAS-Net (Ours)}} & {2-Class} & \textbf{82.34} & \textbf{82.32} \\
     \ & {3-Class} & \textbf{63.10} & -\\
      \midrule
    \bottomrule
  \end{tabular}
  }
  \end{center}
\end{table}

\begin{table}[]
\color{black}
\begin{center}
  \caption{Emotion recognition performance  $\left(\%\right) $ on DEAP dataset.}
  \label{tab:DEAPCompare}
  \scalebox{0.70}{
  \begin{tabular}{lcccccccc}
   \midrule
    \toprule
     \ \multirow{2}{*}{Methods} & \multicolumn{2}{c}{Valence} & \multicolumn{2}{c}{Arousal} & \multicolumn{2}{c}{Dominance} & \multicolumn{2}{c}{Liking}\\
     \ & $P_{acc}$ & $P_f$ & $P_{acc}$ & $P_f$ & $P_{acc}$ & $P_f$ & $P_{acc}$ & $P_f$\\
     \midrule
       \ \textbf{\textit{Supervised}} & \multicolumn{4}{l}{\textbf{\textit{Subject-Dependent}}} & \multicolumn{4}{c}{\textbf{\textit{K-fold CV}}}\\
     \midrule
    \ Li \etal \cite{li2015eeg} & 58.40&-& 	64.20&-& 	65.80&-& 	66.90&-\\
    \ Chen \etal \cite{chen2015electroencephalogram} &67.89&67.83&	69.09&68.96&	-&-&	-&-\\
    \midrule
     \ \textbf{\textit{Supervised}} & \multicolumn{4}{l}{\textbf{\textit{Subject-Dependent}}} & \multicolumn{4}{c}{\textbf{\textit{Video-level LOOCV}}}\\
     \midrule
    \ Koelstra \etal \cite{koelstra2011deap} &57.60&56.30&	62.00&58.30&	-&-&	55.40&50.20 \\
    \ DT-CWPT \cite{naser2013recognition} & 64.30&-&	66.20&-&	68.90&-&	70.20&- \\
    \ EMD \cite{zhuang2017emotion}& 69.10&-&	71.99&-&	-&-&	-&-\\
    \midrule
     \ \textbf{\textit{Supervised}} & \multicolumn{4}{l}{\textbf{\textit{Subject-Independent}}} & \multicolumn{4}{c}{\textbf{\textit{K-fold CV}}}\\
     \midrule
    \ Torres-Valencia \etal \cite{torres2014comparative}& 58.75&-&	55.00&-&	-&-&	-&-\\
    \ Atkinson and Campos \cite{atkinson2016improving}& 73.14&-&	73.06&-&	-&-&	-&-\\
    \ Liu \etal \cite{liu2016emotion} &69.90&-&	71.20&-&	-&-&	-&-\\
    \midrule
      \ \textbf{\textit{Supervised}} & \multicolumn{4}{l}{\textbf{\textit{Subject-Independent}}} & \multicolumn{4}{c}{\textbf{\textit{Subject-level LOOCV}}}\\
      \midrule
    \ Shahnaz \etal \cite{shahnaz2016emotion} &64.71	&74.94&66.51&76.68&	66.88&76.67&	70.52&81.94\\
    \ DGCNN \cite{song2018eeg} &59.29	&-&61.10&-&	-	&-&-&-\\
    \ ATDD-LSTM \cite{du2020efficient} &69.06&-&	72.97&-&	-&-&	-&-\\
    \midrule
    \ \textbf{\textit{Unsupervised}} & \multicolumn{4}{l}{\textbf{\textit{Subject-Dependent}}} & \multicolumn{4}{c}{\textbf{\textit{Video-level LOOCV}}}\\
    \midrule
    \ Liang \etal \cite{liang2019unsupervised} &56.25 &61.25&	62.34 &60.44&	64.22 &64.80&	66.09&77.52\\
    \midrule
    \ \textbf{\textit{Unsupervised}} & \multicolumn{4}{l}{\textbf{\textit{Subject-Independent}}} & \multicolumn{4}{c}{\textbf{\textit{Subject-level LOOCV}}}\\
    \midrule
    \ Liang \etal \cite{liang2019unsupervised} &54.30 & 53.45&	55.55 & 52.77 &	57.03 & 52.84 &58.91& 59.99\\
    \ Baseline \cite{liang2021eegfusenet} &56.44 & 70.83&	58.55 & 72.00 &	61.71 & 74.32 &65.89& 78.46\\
    \ \textbf{TAS-Net (Ours)} & \textbf{57.84}	&\textbf{71.80}&\textbf{60.51}&\textbf{72.64}&	\textbf{63.17}&\textbf{75.32}&	\textbf{67.32}&\textbf{79.74}\\
     \midrule
    \bottomrule
  \end{tabular}
  }
  \end{center}
\end{table}

\begin{table}[]
\color{black}
\begin{center}
  \caption{Emotion recognition performance $\left(\%\right) $ on MAHNOB-HCI dataset.} 
  \label{tab:HCIcompare}
  \scalebox{0.77}{
  \begin{tabular}{lcccccccc}
   \midrule
    \toprule
     \ \multirow{2}{*}{Methods} &\multicolumn{2}{c}{Valence} & \multicolumn{2}{c}{Arousal} & \multicolumn{2}{c}{Dominance} & \multicolumn{2}{c}{Predictability} \\
     \ & $P_{acc}$ & $P_f$ & $P_{acc}$ & $P_f$ & $P_{acc}$ & $P_f$ & $P_{acc}$ & $P_f$ \\
     \midrule
        \ \textbf{\textit{Supervised}} & \multicolumn{4}{l}{\textbf{\textit{Subject-Dependent}}} & \multicolumn{4}{c}{\textbf{\textit{Video-level LOOCV}}}\\
     \midrule
         \ Zhu \etal \cite{zhu2014emotion} & 55.72 & 51.44&60.23 &57.77 & - &-&-&- \\
         \midrule
        \ \textbf{\textit{Supervised}} & \multicolumn{4}{l}{\textbf{\textit{Subject-Independent}}} & \multicolumn{4}{c}{\textbf{\textit{Subject-level LOOCV}}}\\
        \midrule
    \ Soleymani \etal \cite{soleymani2011multimodal} & 57.00 & 56.00 &52.40 &42.00& - &-&-&- \\
    \ Huang \etal \cite{huang2016multi} & 62.13 & - & 61.80 &-& - &-&-&-\\
    \ LRFS \cite{yin2020locally} & 69.93 & 71.22 & 67.43 &68.58 & - &-&-&- \\ 
    \midrule
     \ \textbf{\textit{Unsupervised}} & \multicolumn{4}{l}{\textbf{\textit{Subject-Independent}}} & \multicolumn{4}{c}{\textbf{\textit{Subject-level LOOCV}}}\\
    \midrule
    \ {Baseline} & 60.64 & 72.18 & 62.06	 & 62.05 & 67.08	 & 76.65 & 74.63	& 83.61 \\
    \ \textbf{TAS-Net(Ours)} & \textbf{62.51}	&\textbf{74.64}&\textbf{64.69}& 60.60&	\textbf{69.84}&\textbf{79.34}&	\textbf{75.11}&\textbf{84.10}\\
     \midrule
    \bottomrule
  \end{tabular}
  }
  \end{center}
\end{table}

\begin{figure*}
\begin{center}
\includegraphics[width=1\textwidth]{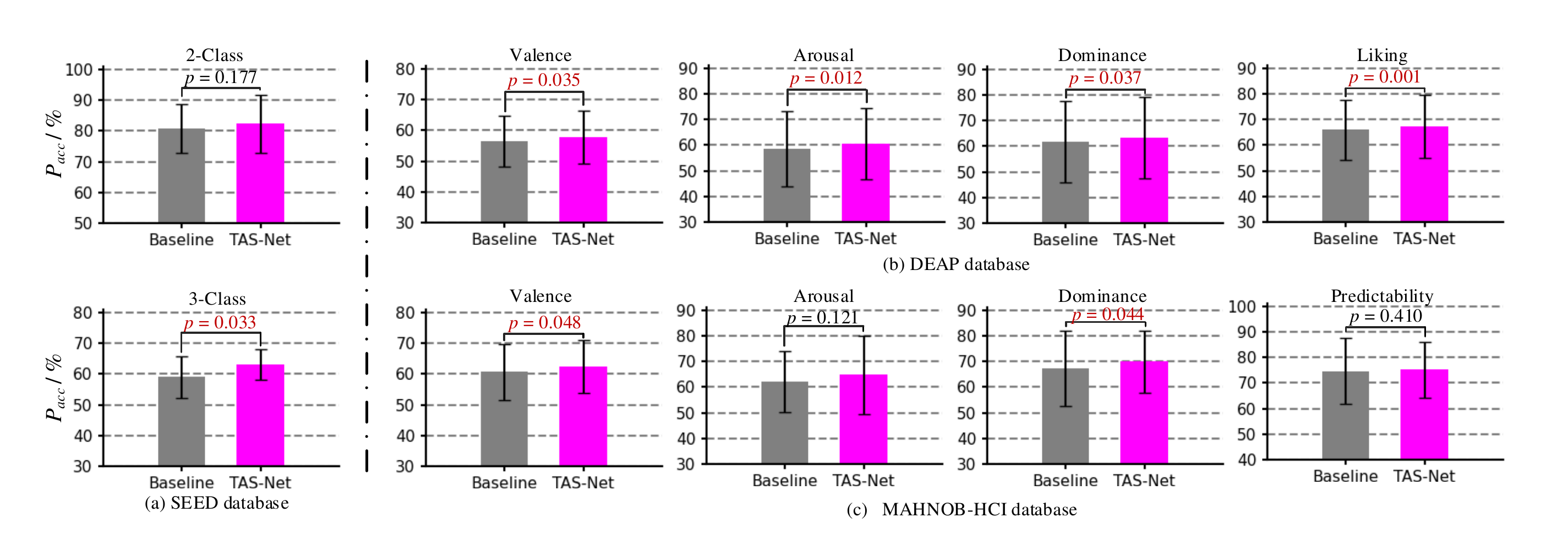}
\end{center}
\caption{The obtained $ p $-values of the paired t-test between the proposed TAS-Net and the baseline method in terms of $ P_{acc} $.}
\label{fig:PValue}
\end{figure*}

\subsection{Evaluation on SEED dataset}
We compare the proposed TAS-Net with the state-of-the-art emotion recognition methods. Table \ref{tab:seedCompare} reports the comparison results, in which the adopted learning methods and cross-validation protocols are clearly stated. In general, compared to the unsupervised learning-based methods, the supervised learning-based methods would achieve better emotion recognition performance, as both data and label information are involved in the model training process. The TAS-Net performance ($P_{acc}$) on the two-class (positive and negative) classification and three-class (positive, neutral, and negative) classification are 82.34\% and 63.10\%, respectively. Compared to the baseline method, the proposed method enhances the average accuracy of the above two classification tasks by 1.87\% and 6.84\%, respectively. The performance difference between the two methods is further evaluated by the paired t-test, and the obtained $ p $-values are reported in Fig. 4(a), which shows that the proposed method has significant differences ($p < 0.05$) compared to the baseline method. On the other hand, our method and the supervised transfer learning method (TCA \cite{pan2010domain}) achieve similar performance. In general, the results shown in Table 1 and Fig. 4(a) prove that the proposed method can perform more accurate unsupervised emotion recognition on the SEED dataset.

\subsection{Evaluation on DEAP dataset}
Similar to the existing studies, a binary classification task is conducted to evaluate the emotion recognition performance of each emotion dimension on the DEAP dataset. Table \ref{tab:DEAPCompare} reported the corresponding results in terms of emotion recognition accuracy $P_{acc}$ and F1-Score ($P_f$). The results show our proposed method achieves a comparable performance compared to the other supervised methods, where the recognition accuracies ($P_{acc}$) of valence, arousal, dominance, and predictability are 57.84$\%$, 60.51$\%$, 63.17$\%$ and 67.32$\%$ and the corresponding F1-Score ($P_{f}$) values are 71.80$\%$, 72.64$\%$, 75.32$\%$, and 79.74$\%$. Compared with the baseline method, our method improved the average accuracy of the four emotion dimensions by 2.48\%, 3.35\%, 2.37\%, and 2.17\%, with significant difference ($p < 0.05$) as shown in Fig. 4 (b). As a result, our method can be regarded as significantly better than the baseline method on the DEAP dataset. Also, compared with the other existing unsupervised EEG decoding method \cite{liang2019unsupervised}, our method also achieves significant performance gains on all emotion dimensions.

\begin{figure*}
\begin{center}
\includegraphics[width=1\textwidth]{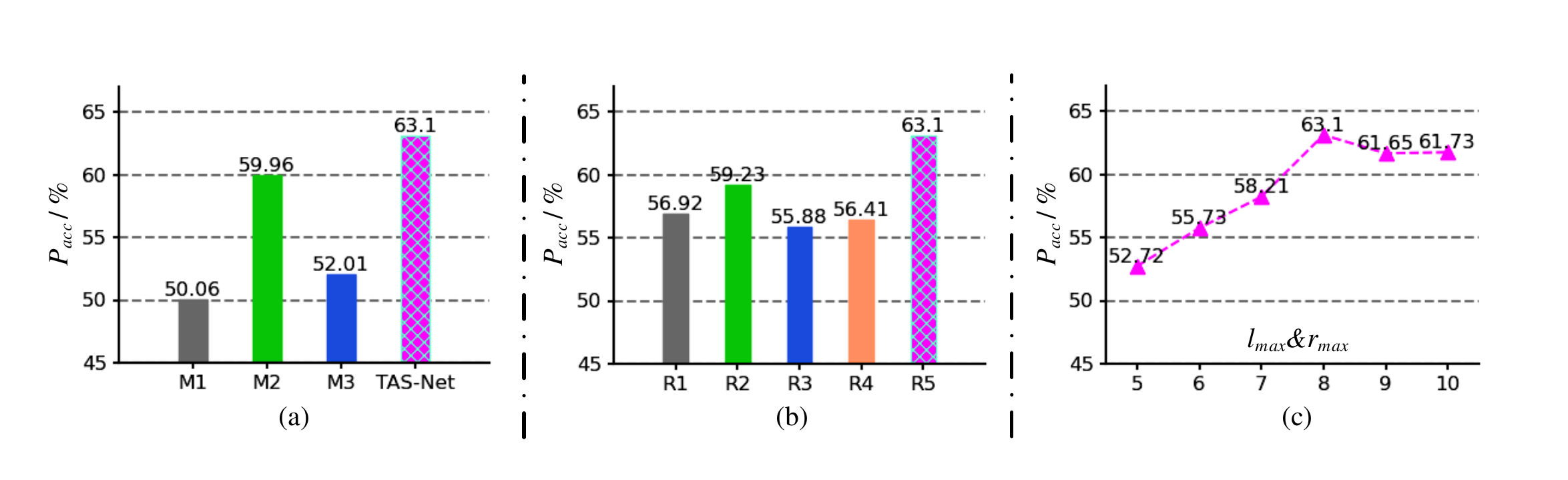}
\end{center}
\caption{Emotion recognition performance (\%) under different methods and parameters comparison.}
\label{fig:Ablation}
\end{figure*}

\begin{table}[]
\color{black}
\begin{center}
  \caption{Emotion recognition performance (\%) with the state-of-the-art clustering methods using subject-independent LOOCV strategy on SEED dataset, under the conditions without (w/o) and with (w) the proposed time-aware sampling method.}
  \label{tab:clusterModelSEED}
  \scalebox{0.9}{
  \begin{tabular}{lccccccc}
  \midrule
    \toprule
    \ \multirow{2}{*}{Methods} &\multirow{2}{*}{Sampling} & \multicolumn{3}{c}{Two-Class} & \multicolumn{2}{c}{Three-Class} \\
    \ & & $P_{acc}$ & $P_f$ & NMI & $P_{acc}$ & NMI \\
    \midrule
    \ \multirow{2}{*}{Simple graph} &w/o & 64.16 &64.15 &0.1728& 35.82 &0.0018\\
     \ &w & 65.96 &66.22 &0.1896 & 56.16 &0.2800\\
     \midrule
    \ \multirow{2}{*}{PCA+k-means} &w/o & 56.12 & 65.27&0.0658& 40.09 &0.0686\\
    \ &w & 60.64 &61.23 &0.0879 & 58.22 &0.2868\\
    \midrule
    \ \multirow{2}{*}{KNN} &w/o & 60.87 &39.18 &0.1139& 42.04 &0.1132\\
    \ &w & 79.72 & 80.04 &0.4237& 52.99 &0.2504\\
    \midrule
    \ \multirow{2}{*}{RCC \cite{shah2017robust}} &w/o & 55.52& 66.24&0.0601& 38.27 &0.0752\\
    \ &w & 58.95 & 69.38 &0.0869& 42.56 &0.1276\\
    \midrule
    \ \multirow{2}{*}{AGDL \cite{zhang2012graph}} &w/o & 51.09 &67.63&0& 34.47 &0\\
     \ &w & 53.45 &67.43 &0.0481& 35.89 &0.0399\\
     \midrule
    \ \multirow{2}{*}{Hypergraph} &w/o &80.83 &82.03 &0.4381 &59.06 &0.3569\\
    \ &w & 82.34 &82.32 &0.4135& 63.10 &0.3750\\
    \midrule
    \bottomrule
  \end{tabular}
  }
  \end{center}
\end{table}

\subsection{Evaluation on MAHNOB-HCI Dataset}
Similarly, a binary classification task is conducted on the MAHNOB-HCI dataset to evaluate the emotion recognition performance of each emotion dimension. As shown in Table \ref{tab:HCIcompare}, our proposed method achieves a comparable performance compared to the other supervised methods, where the recognition accuracies ($P_{acc}$) of valence, arousal, dominance, and predictability are 62.51$\%$, 64.69$\%$, 69.84$\%$ and 75.11$\%$ and the corresponding F1-Score ($P_{f}$) values are 74.64$\%$, 60.60$\%$, 79.34$\%$, and 84.10$\%$. Compared with the baseline approach, our method improved the average accuracy of the above four affective dimensions by 3.08\%, 4.24\%, 4.11\%, and 0.64\%, respectively. The corresponding statistical difference analysis is reported in Fig. 4 (c). It shows, compared with the baseline method, our method achieves significant performance improvement on two emotional dimensions (valence and dominance), while achieving similar performance on the other two emotional dimensions (arousal and predictability).

\section{Discussions} 
\label{sec:discussion}
To fully investigate the effectiveness and efficiency of the proposed TAS-Net, we conduct multiple sets of ablation experiments, including comparing different components, evaluating the defined rewards, evaluating the existing clustering methods with or without TAS-Net, and analyzing the parameter effect. Further, to quantitatively evaluate the key emotion fragment detection performance using TAS-Net, we introduce temporal intersection over union (tIoU) \cite{zhang2022eeg} to calculate the recall rate of samplings. In the performance comparison with different clustering methods, we add normalized mutual information (NMI) as another performance metric to check the corresponding clustering quality. Due to space limitations, all experimental analyses in this section are based on the SEED dataset.

\subsection{Ablation Study}
We conduct ablation experiments to evaluate the contribution of each component in the proposed TAS-Net. The compared methods are: (i) M1: the agent is a FC layer, which produces the action probability. (ii) M2: the agent is a local time-based GCN topped with a FC layer; (iii) M3: the agent is a global time-based BiGRU topped with a FC layer \cite{zhou2018deep}. Fig. 5(a) shows the obtained accuracies when different methods are used for unsupervised emotion recognition. The best performance is achieved by seamlessly combining local time-based GCN and global time-based BiGRU (the proposed TAS-Net). Compared with M1, M2 and M3, our method improves the average accuracy $ P_{acc} $ by 26.05$\%$, 5.24$\%$, and 21.32$\%$, respectively. Therefore, we believe that it is effective to model the time-aware agent by considering that human emotion has not only ”short-term continuity” but also ”long-term similarity”.

\subsection{Effectiveness of Rewards}
To explore the optimization effect of reward functions in the proposed TAS-Net, we set different reward functions for comparative experiments. Here, $\mathcal{R}_{rep}$ is to measure how well these selected key emotion fragments can represent the whole trial's emotional information. Considering the signals evoked by the same emotion should be similar, $\mathcal{R}_{sim}$ is calculated to cater to the "long-term similarity" of emotions. $\mathcal{R}_{div}$ is another commonly used reward function to measure the difference between the selected key emotion fragments in the feature space to evaluate the degree of diversity of the generated summary. Fig. 5(b) shows the obtained unsupervised emotion recognition accuracies using different reward functions. R1, R2, R3, R4 and R5 represent $\mathcal{R}_{rep}$, $\mathcal{R}_{sim}$, $\mathcal{R}_{div}$, $\mathcal{R}_{rep} + \mathcal{R}_{div}$ and $\mathcal{R}_{rep} + \mathcal{R}_{sim}$, respectively. It is found that the reward function of $\mathcal{R}_{sim}$ achieves better performance than $\mathcal{R}_{div}$, which indicates that the similarity reward function is more suitable for the key emotion fragment detection task. Compared with the reward function $\mathcal{R}_{rep}$, when the reward function is composed of $\mathcal{R}_{rep}$ and $\mathcal{R}_{div}$, the recognition performance decreases, while when the reward function is composed of $\mathcal{R}_{rep}$ and $\mathcal{R}_{sim}$, the recognition performance increases, which proves that the same emotion has similarity in different time periods.

\begin{table*}[ht]
\color{black}
\begin{center}
  \caption{The recall $\left(\%\right) $ of emotion localization task on SEED dataset. Note that our method is based on subject-independent validation strategy, while the other two comparison methods are based on subject-dependent validation strategy.} 
  \label{tab:EL-subIndependent}
  \scalebox{0.95}{
  \begin{tabular}{lcccccccccccccccc}
   \midrule
    \toprule
     \  \multirow{2}{*}{Subject ID} &\multicolumn{13}{c}{Stimulus ID} & \multicolumn{3}{c}{Average} \\
     \ &  $\#1$ & $\#2$ & $\#3$ & $\#4$ & $\#5$ & $\#6$ & $\#7$ & $\#8$ & $\#9$ & $\#10$ & $\#11$ & $\#12$ & $\#13$ & Ours & AsI-AEIR\cite{petrantonakis2012adaptive} & Zhang \textit{et al.} \cite{zhang2022eeg}\\
     \midrule
    \ Subject1 & 33 & 50 & 0 & 67 &100 & 67 & 60 & 50 & 50 & 17 & 25 & 50 & 60 & 48 & 40 & 71\\
    \ Subject2 & 33 & 75 & 60 & 50 &100 & 17 & 40 & 25 & 75 & 17 & 75 & 50 & 20 & 49 & 42 & 77\\
    \ Subject3 & 67 & 25 & 80 & 50 &100 & 67 & 40 & 25 & 75 & 33 & 25 & 25 & 60 & 52 & 18 & 76\\
    \ Subject4 & 0 & 50 & 60 & 33 &67 & 17 & 40 & 75 & 50 & 50 & 75 & 25 & 60 & 46 & 45 & 81\\
    \ Subject5 & 100 & 75 & 20 & 33 &100 & 67 & 60 & 50 & 25 & 0 & 50 & 50 & 10 & 52 & 40 & 69\\
    \ Subject6 & 0 & 75	& 60 & 33 & 67	& 50 &	60 & 75 &	25	&33	&25	&75	&40	&48 & 58 & 75\\
    \ Subject7 & 33	&50	&40&	33&	33&	33&	20&	0	&75&	50&	75&	75&	60&	44 & 43 & 71\\
    \ Subject8 & 67	&50&	40&	33	&100&	33&	40&	75&	25&	33&	75&	25&	60&	50 & 76 & 75\\
    \ Subject9 & 67&	50&	80&	33&	100&	50&	60&	50&	0&	33&	25&	50&	20&	48 & 51 & 70\\
    \ Subject10 & 33	&50	&60	&33	&67	&17	&40	&50	&50	&33	&25	&75	&80	&47 & 69 & 72\\
    \ Subject11 & 100&	25	&60	&33	&67	&50	&60	&25	&75	&50	&50	&75	&100	&59 & 73 & 83\\
    \ Subject12 & 67&	50&	80&	67&	67&	50&	80&	0	&25	&50	&75	&25	&20	&50 & 62&72\\
    \ Subject13 & 33&	75&	80&	17&	67&	67&	40&	100&	50&	33&	25&	25&	40&	50 & 39 & 85\\
    \ Subject14 & 100	&25	&20	&33	&67	&67	&60	&50	&75	&50	&50	&50	&40	&53 & 57 & 78\\
   \ Subject15 &67	&100	&20	&33	&100	&17	&60	&50	&50	&33	&25	&50	&60	&51 & 43 & 73\\
    \ All Subjects & 53&	55&	51&	39&	80&	45&	51	&47	&48	&34	&47	&48&	51&	\textbf{50} & \textbf{50} & \textbf{75}\\
    \midrule
    \bottomrule
  \end{tabular}
  }
  \end{center}
\end{table*}

\begin{table*}[ht]
\color{black}
\begin{center}
  \caption{The recall $\left(\%\right) $ of emotion localization task on SEED dataset. Note that all three methods are based on subject-dependent validation strategy.} 
  \label{tab:EL-subDependent}
  \scalebox{0.95}{
  \begin{tabular}{lcccccccccccccccc}
   \midrule
    \toprule
     \ \multirow{2}{*}{Subject ID} &\multicolumn{13}{c}{Stimulus ID} & \multicolumn{3}{c}{Average} \\
     \ & $\#1$ & $\#2$ & $\#3$ & $\#4$ & $\#5$ & $\#6$ & $\#7$ & $\#8$ & $\#9$ & $\#10$ & $\#11$ & $\#12$ & $\#13$ & Ours & AsI-AEIR\cite{petrantonakis2012adaptive} & Zhang \textit{et al.} \cite{zhang2022eeg} \\
     \midrule
    \ Subject1 & 100 & 75  & 80 & 67 &100 & 83 & 80 & 100  & 75 & 83 & 75 & 75  & 80 & 83  & 40 & 71\\
    \ Subject2 & 100 & 100  & 80 & 67 &100 & 67 & 60 & 100  & 100 & 83 & 75 & 75  & 60 & 82  & 42 & 77\\
    \ Subject3 & 100 & 75  & 80 & 67 &100 & 83 & 100 &75  & 75 & 67 & 75 & 75  & 80 & 81  & 18 & 76\\
    \ Subject4 & 67 & 100  & 80 & 67 &100 & 100 & 80 & 75  & 75 & 67 & 100 & 75  & 80 & 82  & 45 & 81\\
    \ Subject5 & 67 & 75  & 80 & 83 &100 & 83 & 80 & 100  & 75 & 67 & 75 & 75  & 80 & 80  & 40 & 69\\
    \ Subject6 &100&	75&	80&	67&	100&	67	&80	&75&	75&	83&	75&	75&	80&	79  & 58 & 75\\
    \ Subject7 & 100&	75&	80&	83&	100&	83&	80&	75&	100&	83&	75&	100&	80&	86  & 43 & 71\\
    \ Subject8 & 100&	75&	80&	67&	100&	67&	80&	100&	75&	67&	75&	75&	80&	80  & 76 & 75\\
    \ Subject9 & 67	&75	&80&	67&	100&	83&	60&	75&	75&	83&	75&	100&	80&	78  & 51 & 70\\
    \ Subject10 & 100	&75	&80	&67	&100	&50	&40	&75	&75	&67	&75	&75	&80	&74  & 69 & 72\\
    \ Subject11 & 67&	100	&80	&67	&100	&67	&80	&75	&75	&83	&100	&75	&60	&79  & 73 & 83\\
    \ Subject12 & 100&	75&	80&	67&	100&	83&	60&	75	&100	&83	&75	&75	&80	&81 & 62&72\\
    \ Subject13 & 67&	75&	80&	67&	100&	83&	80&	75&	75&	67&	100&	75&	80&	79  & 39 & 85\\
    \ Subject14 & 100	&75	&80	&83	&100	&83	&80	&100	&100	&67	&75	&75	&80	&84  & 57 & 78\\
    \ Subject15 &100	&75	&100	&83	&100	&67	&80	&75	&100	&83	&75	&100	&80	&86  & 43 & 73\\
    \ All Subjects & 89	&80&	81&	71&	100&	77&	75&	83&	83&	76&	80&	80&	77	&	\textbf{81}  & \textbf{50} & \textbf{75}\\
    \midrule
    \bottomrule
  \end{tabular}
  }
  \end{center}
\end{table*}

\subsection{Performance Comparison Using Different Clustering Methods}
The purpose of this study is to design a sampling model that can efficiently localize short-term continuous fragments with correlated emotions in EEG time series, so as to achieve better unsupervised emotion recognition performance. Theoretically, the sampling set can effectively improve the emotion recognition performance of different clustering methods. We employ six decoding models for this comparative experiment, including a simple graph based method, principal component analysis (PCA) and K-means clustering method (PCA+Kmeans), K-nearest neighbors (KNN) algorithm, robust continuous clustering method (RCC), directed graph based agglomerative algorithm (AGDL), and hypergraph algorithm. The simple graph completes unsupervised clustering by measuring pair-wise relationships. K-means is one of the outstanding representatives of unsupervised clustering, and KNN is one of the simplest classification algorithms in supervised clustering. RCC is a fast, simple, and efficient high-dimensional clustering algorithm that achieves decoding of highly mixed clusters by optimizing a clear global objective. AGDL is a graph-based agglomerative algorithm, which quantifies the stability of clustering results by computing the product of average indegree and average outdegree. Table 4 reports the unsupervised recognition accuracies of different clustering methods without or with the time-aware sampling method. In the case of using the time-aware sampling method, significant performance improvement can be observed in all the above clustering methods, which shows the effectiveness of the proposed time sampling method in unsupervised based emotion recognition using EEG signals.

\subsection{Parameter Analysis}
We further quantitatively evaluate the effectiveness of the proposed TAS-Net under different parameters. There are two important hyper-parameters in our method: the number $ K $ of selected fragments and the maximum absolute offset to the left and right parameter $ l_{max} \& r_{max} $. The setting of $ K $ can affect the recall rate of sampling and the accuracy of emotion recognition, while $ l_{max} $\&$ r_{max} $ determines the maximum length of a selected fragment. Following \cite{zhang2022eeg}, we set $ K $ to 10 and set $ l_{max} $\&$ r_{max} $ to 5, 6, 7, 8, 9, 10. As shown in Fig. 5(c), our method achieves the best performance when $ l_{max} $\&$ r_{max} $ is 8, which means that the max length of a fragment with coherent stronger emotion is 16. According to the experimental results, we clearly find that our method is sensitive to this parameter when $ l_{max} $\&$ r_{max} $ is less than 8, while the results obtained by our method are relatively stable when $ l_{max} $\&$ r_{max} $ is greater than 8.

\begin{figure*}
\begin{center}
\includegraphics[width=1\textwidth]{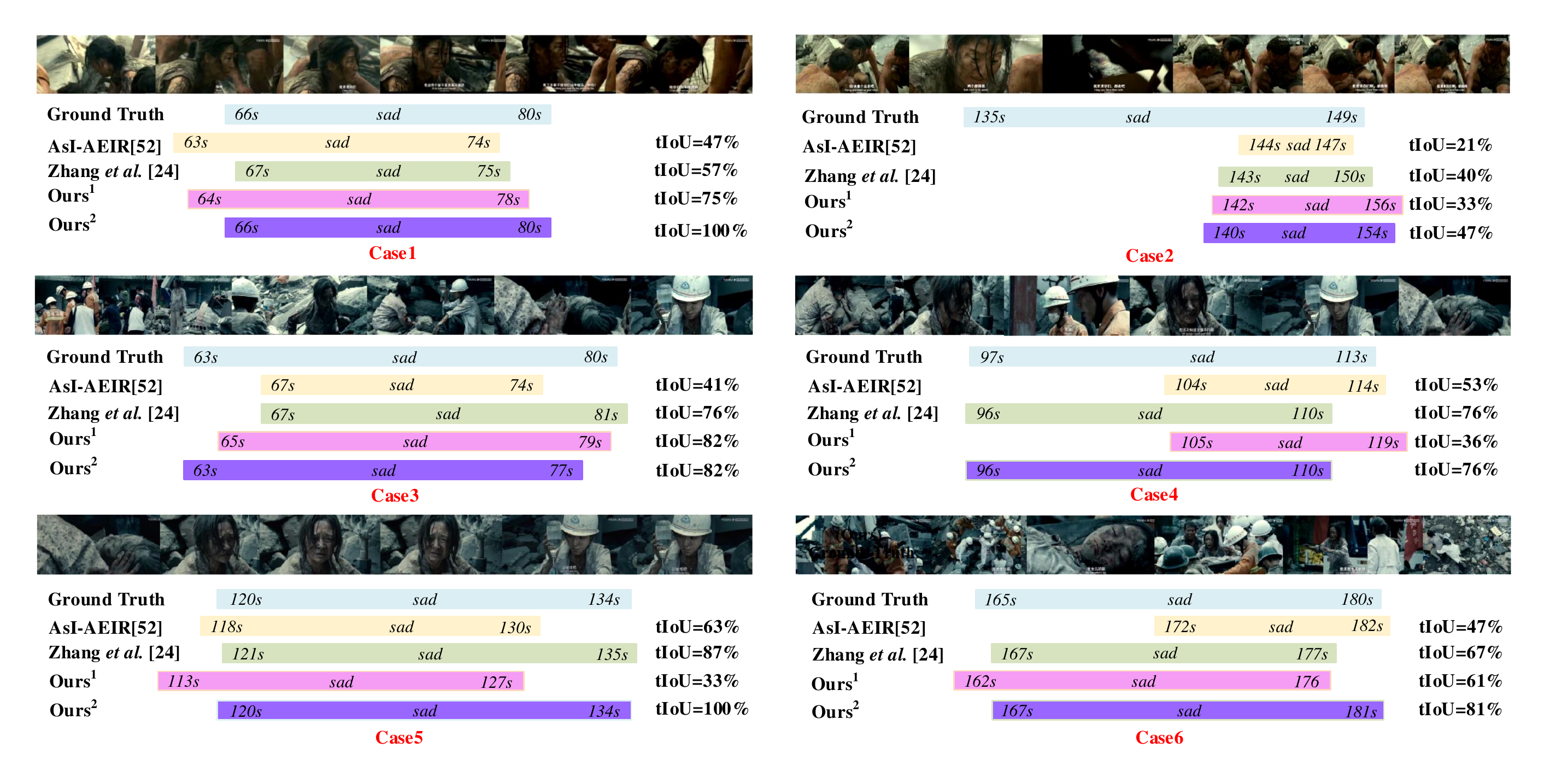}
\end{center}
\caption{Visualization of selected key fragments in different videos. Ours$^1$ and Ours$^2$ represent the results obtained by our method based on subject-independent LOOCV and subject-dependent LOOCV, respectively.}
\label{fig:sampling}
\end{figure*}

\subsection{Performance on Emotion Localization}
Recently, Zhang \textit{et al.}\cite{zhang2022eeg} proposed a supervised learning based hierarchical self-attention network on SEED dataset, which incorporates emotion localization task for improving emotion recognition performance. Similarly, we also extend our proposed TAS-Net on emotion localization application under an unsupervised learning manner. In the emotion localization task, EEG signals corresponding to each video stimulus were segmented, and then twenty people were asked to score each fragment according to the corresponding emotion to obtain the ground truth labels. In addition, the recall is used as the evaluation metric for this task, and the threshold of tIoU (the overlap rate in the timeline between the ground truth fragments and the top-$ X $ selected fragments $f_{x}$) is set to 0.5. Note here that, for cross-comparison with \cite{zhang2022eeg}, both subject-independent LOOCV and subject-dependent LOOCV are used for emotion localization in our method. Tables 5 and 6 report the recall rates of emotion localization obtained by our method using subject-independent LOOCV and subject-dependent LOOCV, respectively. Compared with the other two emotion localization methods that employ subject-dependent LOOCV, our method achieves the highest performance. When subject-independent LOOCV is used (which is more difficult than subject-dependent LOOCV due to the individual differences), our model's subject-independent LOOCV results are comparable with AsI-AEIR \cite{petrantonakis2012adaptive}'s subject-dependent LOOCV results. In other words, our model is more effective and efficient in emotion localization under both subject-dependent LOOCV and subject-independent LOOCV. For a more intuitive comparison, we also visualize the emotion localization results when different methods are used. As shown in Fig. 6, the ground truth labels with the localization results and tIoU values are reported, which shows the possibility to detect key emotion fragments even when label information is missing.

\section{Conclusion}
This paper develops an unsupervised time-aware sampling network (TAS-Net) based on deep reinforcement learning to adaptively perceive the location of strong emotions in continuously collected EEG signals, thereby improving the performance of unsupervised emotion recognition. To avoid a flat importance score for each sample-level feature in the EEG time series, the proposed TAS-Net accomplishes more efficient key emotion fragment detection by seamlessly combining local time-based GCN and a global time-based BiGRU. To the best of our knowledge, we are the first to utilize deep reinforcement learning to complete key emotion fragment detection in EEG signals for more accurate unsupervised emotion recognition. Experiments on three challenging public datasets show that our TAS-Net achieves excellent performance. The performance in detecting key emotion fragments from the continuously collected EEG signals has been verified, and the superiority of emotion recognition has been demonstrated. Since the label information is not required by the proposed new method, TAS-Net is a fully unsupervised method that could be easily extended to other EEG-related applications and be utilized to automatically extract essential information from time-series EEG data.


%




\section{Conflicts of Interest}
The authors declare that they have no conflicts of interest.

\section{Acknowledgment}
This work was supported in part by the National Natural Science Foundation of China under Grant 82272114, 62276169, 61906122, and 62071310, in part by Shenzhen-Hong Kong Institute of Brain Science-Shenzhen Fundamental Research Institutions (2022SHIBS0003), in part by Shenzhen Science and Technology Research and Development Fund for Sustainable Development Project (No.KCXFZ20201221173613036).

\ifCLASSOPTIONcaptionsoff
  \newpage
\fi

\bibliographystyle{IEEEtran}
\bibliography{references}

\begin{thebibliography}{10}
\providecommand{\url}[1]{#1}
\csname url@samestyle\endcsname
\providecommand{\newblock}{\relax}
\providecommand{\bibinfo}[2]{#2}
\providecommand{\BIBentrySTDinterwordspacing}{\spaceskip=0pt\relax}
\providecommand{\BIBentryALTinterwordstretchfactor}{4}
\providecommand{\BIBentryALTinterwordspacing}{\spaceskip=\fontdimen2\font plus
\BIBentryALTinterwordstretchfactor\fontdimen3\font minus
  \fontdimen4\font\relax}
\providecommand{\BIBforeignlanguage}[2]{{%
\expandafter\ifx\csname l@#1\endcsname\relax
\typeout{** WARNING: IEEEtran.bst: No hyphenation pattern has been}%
\typeout{** loaded for the language `#1'. Using the pattern for}%
\typeout{** the default language instead.}%
\else
\language=\csname l@#1\endcsname
\fi
#2}}
\providecommand{\BIBdecl}{\relax}
\BIBdecl

\bibitem{cowen2017self}
A.~S. Cowen and D.~Keltner, ``Self-report captures 27 distinct categories of
  emotion bridged by continuous gradients,'' \emph{Proceedings of the national
  academy of sciences}, vol. 114, no.~38, pp. E7900--E7909, 2017.

\bibitem{horikawa2020neural}
T.~Horikawa, A.~S. Cowen, D.~Keltner, and Y.~Kamitani, ``The neural
  representation of visually evoked emotion is high-dimensional, categorical,
  and distributed across transmodal brain regions,'' \emph{Iscience}, vol.~23,
  no.~5, p. 101060, 2020.

\bibitem{huang2014novel}
Y.-J. Huang, C.-Y. Wu, A.~M.-K. Wong, and B.-S. Lin, ``Novel active comb-shaped
  dry electrode for eeg measurement in hairy site,'' \emph{IEEE Transactions on
  Biomedical Engineering}, vol.~62, no.~1, pp. 256--263, 2014.

\bibitem{liu2017real}
Y.-J. Liu, M.~Yu, G.~Zhao, J.~Song, Y.~Ge, and Y.~Shi, ``Real-time
  movie-induced discrete emotion recognition from eeg signals,'' \emph{IEEE
  Transactions on Affective Computing}, vol.~9, no.~4, pp. 550--562, 2017.

\bibitem{song2018eeg}
T.~Song, W.~Zheng, P.~Song, and Z.~Cui, ``Eeg emotion recognition using
  dynamical graph convolutional neural networks,'' \emph{IEEE Transactions on
  Affective Computing}, vol.~11, no.~3, pp. 532--541, 2018.

\bibitem{li2022eeg}
X.~Li, Y.~Zhang, P.~Tiwari, D.~Song, B.~Hu, M.~Yang, Z.~Zhao, N.~Kumar, and
  P.~Marttinen, ``Eeg based emotion recognition: A tutorial and review,''
  \emph{ACM Computing Surveys (CSUR)}, 2022.

\bibitem{alsolamy2016emotion}
M.~Alsolamy and A.~Fattouh, ``Emotion estimation from eeg signals during
  listening to quran using psd features,'' in \emph{2016 7th International
  Conference on Computer Science and Information Technology (CSIT)}.\hskip 1em
  plus 0.5em minus 0.4em\relax IEEE, 2016, pp. 1--5.

\bibitem{duan2013differential}
R.-N. Duan, J.-Y. Zhu, and B.-L. Lu, ``Differential entropy feature for
  eeg-based emotion classification,'' in \emph{2013 6th International IEEE/EMBS
  Conference on Neural Engineering (NER)}.\hskip 1em plus 0.5em minus
  0.4em\relax IEEE, 2013, pp. 81--84.

\bibitem{duan2012eeg}
R.-N. Duan, X.-W. Wang, and B.-L. Lu, ``Eeg-based emotion recognition in
  listening music by using support vector machine and linear dynamic system,''
  in \emph{International Conference on Neural Information Processing}.\hskip
  1em plus 0.5em minus 0.4em\relax Springer, 2012, pp. 468--475.

\bibitem{lin2010eeg}
Y.-P. Lin, C.-H. Wang, T.-P. Jung, T.-L. Wu, S.-K. Jeng, J.-R. Duann, and J.-H.
  Chen, ``Eeg-based emotion recognition in music listening,'' \emph{IEEE
  Transactions on Biomedical Engineering}, vol.~57, no.~7, pp. 1798--1806,
  2010.

\bibitem{zheng2015investigating}
W.-L. Zheng and B.-L. Lu, ``Investigating critical frequency bands and channels
  for eeg-based emotion recognition with deep neural networks,'' \emph{IEEE
  Transactions on autonomous mental development}, vol.~7, no.~3, pp. 162--175,
  2015.

\bibitem{liu2013real}
Y.~Liu and O.~Sourina, ``Real-time fractal-based valence level recognition from
  eeg,'' in \emph{Transactions on computational science XVIII}.\hskip 1em plus
  0.5em minus 0.4em\relax Springer, 2013, pp. 101--120.

\bibitem{bahari2013eeg}
F.~Bahari and A.~Janghorbani, ``Eeg-based emotion recognition using recurrence
  plot analysis and k nearest neighbor classifier,'' in \emph{2013 20th Iranian
  Conference on Biomedical Engineering (ICBME)}.\hskip 1em plus 0.5em minus
  0.4em\relax IEEE, 2013, pp. 228--233.

\bibitem{atkinson2016improving}
J.~Atkinson and D.~Campos, ``Improving bci-based emotion recognition by
  combining eeg feature selection and kernel classifiers,'' \emph{Expert
  Systems with Applications}, vol.~47, pp. 35--41, 2016.

\bibitem{pan2010domain}
S.~J. Pan, I.~W. Tsang, J.~T. Kwok, and Q.~Yang, ``Domain adaptation via
  transfer component analysis,'' \emph{IEEE transactions on neural networks},
  vol.~22, no.~2, pp. 199--210, 2010.

\bibitem{liu20213dcann}
S.~Liu, X.~Wang, L.~Zhao, B.~Li, W.~Hu, J.~Yu, and Y.~Zhang, ``3dcann: a
  spatio-temporal convolution attention neural network for eeg emotion
  recognition,'' \emph{IEEE Journal of Biomedical and Health Informatics},
  2021.

\bibitem{li2019domain}
J.~Li, S.~Qiu, C.~Du, Y.~Wang, and H.~He, ``Domain adaptation for eeg emotion
  recognition based on latent representation similarity,'' \emph{IEEE
  Transactions on Cognitive and Developmental Systems}, vol.~12, no.~2, pp.
  344--353, 2019.

\bibitem{li2018novel}
Y.~Li, W.~Zheng, Z.~Cui, T.~Zhang, and Y.~Zong, ``A novel neural network model
  based on cerebral hemispheric asymmetry for eeg emotion recognition.'' in
  \emph{IJCAI}, 2018, pp. 1561--1567.

\bibitem{zhong2020eeg}
P.~Zhong, D.~Wang, and C.~Miao, ``Eeg-based emotion recognition using
  regularized graph neural networks,'' \emph{IEEE Transactions on Affective
  Computing}, 2020.

\bibitem{li2019regional}
Y.~Li, W.~Zheng, L.~Wang, Y.~Zong, and Z.~Cui, ``From regional to global brain:
  A novel hierarchical spatial-temporal neural network model for eeg emotion
  recognition,'' \emph{IEEE Transactions on Affective Computing}, 2019.

\bibitem{tao2020eeg}
W.~Tao, C.~Li, R.~Song, J.~Cheng, Y.~Liu, F.~Wan, and X.~Chen, ``Eeg-based
  emotion recognition via channel-wise attention and self attention,''
  \emph{IEEE Transactions on Affective Computing}, 2020.

\bibitem{zheng2016multichannel}
W.~Zheng, ``Multichannel eeg-based emotion recognition via group sparse
  canonical correlation analysis,'' \emph{IEEE Transactions on Cognitive and
  Developmental Systems}, vol.~9, no.~3, pp. 281--290, 2016.

\bibitem{liang2019unsupervised}
Z.~Liang, S.~Oba, and S.~Ishii, ``An unsupervised eeg decoding system for human
  emotion recognition,'' \emph{Neural Networks}, vol. 116, pp. 257--268, 2019.

\bibitem{liang2021eegfusenet}
Z.~Liang, R.~Zhou, L.~Zhang, L.~Li, G.~Huang, Z.~Zhang, and S.~Ishii,
  ``Eegfusenet: Hybrid unsupervised deep feature characterization and fusion
  for high-dimensional eeg with an application to emotion recognition,''
  \emph{IEEE Transactions on Neural Systems and Rehabilitation Engineering},
  vol.~29, pp. 1913--1925, 2021.

\bibitem{lample2017playing}
G.~Lample and D.~S. Chaplot, ``Playing fps games with deep reinforcement
  learning,'' in \emph{Thirty-First AAAI Conference on Artificial
  Intelligence}, 2017.

\bibitem{wang2016does}
F.-Y. Wang, J.~J. Zhang, X.~Zheng, X.~Wang, Y.~Yuan, X.~Dai, J.~Zhang, and
  L.~Yang, ``Where does alphago go: From church-turing thesis to alphago thesis
  and beyond,'' \emph{IEEE/CAA Journal of Automatica Sinica}, vol.~3, no.~2,
  pp. 113--120, 2016.

\bibitem{he2015deep}
J.~He, J.~Chen, X.~He, J.~Gao, L.~Li, L.~Deng, and M.~Ostendorf, ``Deep
  reinforcement learning with a natural language action space,'' \emph{arXiv
  preprint arXiv:1511.04636}, 2015.

\bibitem{yarats2020image}
D.~Yarats, I.~Kostrikov, and R.~Fergus, ``Image augmentation is all you need:
  Regularizing deep reinforcement learning from pixels,'' in
  \emph{International Conference on Learning Representations}, 2020.

\bibitem{zoph2016neural}
B.~Zoph and Q.~V. Le, ``Neural architecture search with reinforcement
  learning,'' \emph{arXiv preprint arXiv:1611.01578}, 2016.

\bibitem{mnih2013playing}
V.~Mnih, K.~Kavukcuoglu, D.~Silver, A.~Graves, I.~Antonoglou, D.~Wierstra, and
  M.~Riedmiller, ``Playing atari with deep reinforcement learning,''
  \emph{arXiv preprint arXiv:1312.5602}, 2013.

\bibitem{zhang2018fuzzy}
D.~Zhang, L.~Yao, S.~Wang, K.~Chen, Z.~Yang, and B.~Benatallah, ``Fuzzy
  integral optimization with deep q-network for eeg-based intention
  recognition,'' in \emph{Pacific-Asia Conference on Knowledge Discovery and
  Data Mining}.\hskip 1em plus 0.5em minus 0.4em\relax Springer, 2018, pp.
  156--168.

\bibitem{li2021eeg}
C.~Li, Z.~Zhang, R.~Song, J.~Cheng, Y.~Liu, and X.~Chen, ``Eeg-based emotion
  recognition via neural architecture search,'' \emph{IEEE Transactions on
  Affective Computing}, 2021.

\bibitem{zhou2018deep}
K.~Zhou, Y.~Qiao, and T.~Xiang, ``Deep reinforcement learning for unsupervised
  video summarization with diversity-representativeness reward,'' in
  \emph{Proceedings of the AAAI Conference on Artificial Intelligence},
  vol.~32, no.~1, 2018.

\bibitem{williams1992simple}
R.~J. Williams, ``Simple statistical gradient-following algorithms for
  connectionist reinforcement learning,'' \emph{Machine learning}, vol.~8,
  no.~3, pp. 229--256, 1992.

\bibitem{zhou2006learning}
D.~Zhou, J.~Huang, and B.~Sch{\"o}lkopf, ``Learning with hypergraphs:
  Clustering, classification, and embedding,'' \emph{Advances in neural
  information processing systems}, vol.~19, 2006.

\bibitem{koelstra2011deap}
S.~Koelstra, C.~Muhl, M.~Soleymani, J.-S. Lee, A.~Yazdani, T.~Ebrahimi, T.~Pun,
  A.~Nijholt, and I.~Patras, ``Deap: A database for emotion analysis; using
  physiological signals,'' \emph{IEEE transactions on affective computing},
  vol.~3, no.~1, pp. 18--31, 2011.

\bibitem{soleymani2011multimodal}
M.~Soleymani, J.~Lichtenauer, T.~Pun, and M.~Pantic, ``A multimodal database
  for affect recognition and implicit tagging,'' \emph{IEEE transactions on
  affective computing}, vol.~3, no.~1, pp. 42--55, 2011.

\bibitem{kwak2002input}
N.~Kwak and C.-H. Choi, ``Input feature selection by mutual information based
  on parzen window,'' \emph{IEEE transactions on pattern analysis and machine
  intelligence}, vol.~24, no.~12, pp. 1667--1671, 2002.

\bibitem{zhang2022eeg}
Y.~Zhang, H.~Liu, D.~Zhang, X.~Chen, T.~Qin, and Q.~Zheng, ``Eeg-based emotion
  recognition with emotion localization via hierarchical self-attention,''
  \emph{IEEE Transactions on Affective Computing}, 2022.

\bibitem{li2015eeg}
X.~Li, P.~Zhang, D.~Song, G.~Yu, Y.~Hou, and B.~Hu, ``Eeg based emotion
  identification using unsupervised deep feature learning,'' 2015.

\bibitem{chen2015electroencephalogram}
J.~Chen, B.~Hu, P.~Moore, X.~Zhang, and X.~Ma, ``Electroencephalogram-based
  emotion assessment system using ontology and data mining techniques,''
  \emph{Applied Soft Computing}, vol.~30, pp. 663--674, 2015.

\bibitem{naser2013recognition}
D.~S. Naser and G.~Saha, ``Recognition of emotions induced by music videos
  using dt-cwpt,'' in \emph{2013 Indian Conference on Medical Informatics and
  Telemedicine (ICMIT)}.\hskip 1em plus 0.5em minus 0.4em\relax IEEE, 2013, pp.
  53--57.

\bibitem{zhuang2017emotion}
N.~Zhuang, Y.~Zeng, L.~Tong, C.~Zhang, H.~Zhang, and B.~Yan, ``Emotion
  recognition from eeg signals using multidimensional information in emd
  domain,'' \emph{BioMed research international}, vol. 2017, 2017.

\bibitem{torres2014comparative}
C.~A. Torres-Valencia, H.~F. Garcia-Arias, M.~A.~A. Lopez, and A.~A.
  Orozco-Guti{\'e}rrez, ``Comparative analysis of physiological signals and
  electroencephalogram (eeg) for multimodal emotion recognition using
  generative models,'' in \emph{2014 XIX Symposium on Image, Signal Processing
  and Artificial Vision}.\hskip 1em plus 0.5em minus 0.4em\relax IEEE, 2014,
  pp. 1--5.

\bibitem{liu2016emotion}
J.~Liu, H.~Meng, A.~Nandi, and M.~Li, ``Emotion detection from eeg
  recordings,'' in \emph{2016 12th international conference on natural
  computation, fuzzy systems and knowledge discovery (ICNC-FSKD)}.\hskip 1em
  plus 0.5em minus 0.4em\relax IEEE, 2016, pp. 1722--1727.

\bibitem{shahnaz2016emotion}
C.~Shahnaz, S.~S. Hasan \emph{et~al.}, ``Emotion recognition based on wavelet
  analysis of empirical mode decomposed eeg signals responsive to music
  videos,'' in \emph{2016 IEEE Region 10 Conference (TENCON)}.\hskip 1em plus
  0.5em minus 0.4em\relax IEEE, 2016, pp. 424--427.

\bibitem{du2020efficient}
X.~Du, C.~Ma, G.~Zhang, J.~Li, Y.-K. Lai, G.~Zhao, X.~Deng, Y.-J. Liu, and
  H.~Wang, ``An efficient lstm network for emotion recognition from
  multichannel eeg signals,'' \emph{IEEE Transactions on Affective Computing},
  2020.

\bibitem{zhu2014emotion}
Y.~Zhu, S.~Wang, and Q.~Ji, ``Emotion recognition from users' eeg signals with
  the help of stimulus videos,'' in \emph{2014 IEEE international conference on
  multimedia and expo (ICME)}.\hskip 1em plus 0.5em minus 0.4em\relax IEEE,
  2014, pp. 1--6.

\bibitem{huang2016multi}
X.~Huang, J.~Kortelainen, G.~Zhao, X.~Li, A.~Moilanen, T.~Sepp{\"a}nen, and
  M.~Pietik{\"a}inen, ``Multi-modal emotion analysis from facial expressions
  and electroencephalogram,'' \emph{Computer Vision and Image Understanding},
  vol. 147, pp. 114--124, 2016.

\bibitem{yin2020locally}
Z.~Yin, L.~Liu, J.~Chen, B.~Zhao, and Y.~Wang, ``Locally robust eeg feature
  selection for individual-independent emotion recognition,'' \emph{Expert
  Systems with Applications}, vol. 162, p. 113768, 2020.

\bibitem{shah2017robust}
S.~A. Shah and V.~Koltun, ``Robust continuous clustering,'' \emph{Proceedings
  of the National Academy of Sciences}, vol. 114, no.~37, pp. 9814--9819, 2017.

\bibitem{zhang2012graph}
W.~Zhang, X.~Wang, D.~Zhao, and X.~Tang, ``Graph degree linkage: Agglomerative
  clustering on a directed graph,'' in \emph{European conference on computer
  vision}.\hskip 1em plus 0.5em minus 0.4em\relax Springer, 2012, pp. 428--441.

\bibitem{petrantonakis2012adaptive}
P.~C. Petrantonakis and L.~J. Hadjileontiadis, ``Adaptive emotional information
  retrieval from eeg signals in the time-frequency domain,'' \emph{IEEE
  Transactions on Signal Processing}, vol.~60, no.~5, pp. 2604--2616, 2012.

\end{thebibliography}

\end{document}